# Brain Tumor Detection Through Diverse CNN Architectures in IoT Healthcare Industries: Fast R-CNN, U-Net, Transfer Learning-Based CNN, and Fully Connected CNN

Mohsen Asghari Ilani and Yaser M. Banad[*]

School of Electrical and Computer Engineering, University of Oklahoma, Norman, Oklahoma, 73019, USA
(*Corresponding author's e-mail: bana@ou.edu)

**Abstract**
Artificial intelligence-powered deep learning methods have significantly advanced the diagnosis of brain tumors in Internet of Things (IoT)-healthcare systems, achieving high accuracy by processing extensive datasets. Brain health is crucial for human life, and accurate diagnosis is vital for effective treatment. Magnetic Resonance Imaging (MRI) provides critical data for diagnosing brain health issues, offering a substantial source of big data for artificial intelligence applications in image classification. In this study, we aimed to classify brain tumors, specifically glioma, meningioma, and pituitary tumors, from MRI images using Region-based Convolutional Neural Network (R-CNN) and UNet architectures. Additionally, we employed Convolutional Neural Networks (CNN) and CNN-based models such as Inception-V3, EfficientNetB4, and VGG19, leveraging transfer learning methods for classification tasks. The models were evaluated using F-score, recall, precision, and accuracy metrics. Our findings revealed that the Fast R-CNN model achieved the highest accuracy at 99%, with an F-score of 98.5%, an Area Under the Curve (AUC) value of 99.5%, a recall of 99.4%, and a precision of 98.5%. The integration of R-CNN, UNet, and transfer learning models plays a pivotal role in the early diagnosis and prompt treatment of brain tumors in IoT healthcare systems, significantly improving patient outcomes. IoT devices, such as wearable health monitors and smart imaging systems, continuously collect and transmit real-time data. This real-time data can be analyzed by AI algorithms to provide immediate and accurate insights into a patient's condition, enabling timely interventions and personalized treatment plans. Moreover, for external cohort cross-dataset validation, among the various fine-tuned EfficientNet models, EfficientNetB2 achieved the highest performance. It demonstrated precision, recall/sensitivity, specificity, F1-score, and accuracy values of 92.11%, 92.11%, 95.96%, 92.02%, and 92.23%, respectively. These results highlight the robustness and reliability of AI models in handling diverse datasets, further emphasizing their potential to enhance brain tumor classification and improve patient care in IoT healthcare systems.

## Introduction

The healthcare industry has been rapidly transformed by technological advances in recent years, with artificial intelligence (AI) technology playing a pivotal role in this transformation. AI, which simulates human-like intelligence, has many applications in medicine, including the fight against brain tumors. Brain tumors are a major public health problem, and accurate diagnosis, treatment, and follow-up are critical. AI has become an important tool for enhancing these processes, offering great potential for early diagnosis and treatment of brain tumors [1].

The integration of the Internet of Things (IoT) with AI has further revolutionized the application of these technologies in healthcare. IoT devices [2-5], such as wearable health monitors and smart imaging systems, continuously collect and transmit vast amounts of data. This real-time data can be analyzed by AI algorithms to provide immediate and accurate insights into a patient's condition. In the context of brain tumor detection, IoT-enabled devices can provide continuous monitoring and rapid diagnostics, allowing for timely interventions and personalized treatment plans [6].

AI-driven models, especially when combined with IoT, have shown remarkable efficacy in brain tumor classification tasks. For instance, advanced models like U-Net and fine-tuned EfficientNets have demonstrated high performance in accurately classifying tumor types, which is crucial for effective treatment planning. The ability to process and analyze large datasets from various sources, facilitated by IoT, enhances the robustness and reliability of these AI models. Consequently, the synergy between AI and IoT not only improves diagnostic accuracy but also supports healthcare professionals in making informed decisions, ultimately leading to better patient outcomes and improved quality of life [7].

In the United States, there are approximately 110 million adolescents and young adults (AYAs) aged 15 to 39. Among them, about 208,620 are living with a primary brain or spinal cord tumor. Although these central nervous system (CNS) tumors are relatively rare, they are the most prevalent cancer in this age group. CNS tumors are also the second leading cause of cancer-related death in AYAs and the leading cause of death for those aged 15 to 24 [8].

As illustrated in *Figure 1*, the location of brain tumors varies by age. Younger AYAs typically develop tumors in CNS areas that form early in development, such as the cerebellum and brainstem. These tumors become less common with age, while tumors of the meninges (accounting for 16% of mortality), the four lobes of the brain, and the pituitary gland become more frequent. Non-malignant pituitary tumors are the most common CNS tumors in the AYA population, particularly for those under 30, representing 36.5% of tumors. The most common malignant tumors are adult-type diffuse gliomas, particularly lower-grade gliomas like astrocytomas and oligodendrogliomas, which make up 3% of all brain tumors. Adults over 40 are more likely to develop higher-grade gliomas, especially glioblastomas, which is why most research focuses on these tumors.

Neuroimaging for brain tumors involves using advanced imaging techniques to visualize and diagnose tumors in the brain. These methods are crucial for accurately identifying the presence, location, size, and characteristics of brain tumors, aiding in the formulation of effective treatment plans and monitoring therapeutic responses. High-resolution imaging techniques, such as X-rays, computed axial tomography (CAT) scans, magnetic resonance imaging (MRI), and positron emission tomography (PET) scans, are commonly used. While X-rays provide valuable insights into bone structures, their limitations in visualizing soft tissues make them less suitable for brain tumor evaluation. MRI offers significant advantages over CT scans, including the ability to detect flowing blood, cryptic vascular malformations, and demyelinating diseases, without the beam-hardening artifacts common in CT. Additionally, MRI provides superior visualization of the posterior fossa and involves no ionizing radiation [9].

MRI uses a strong magnetic field to align protons in tissue water nuclei. Radio Frequency (RF) energy disrupts this alignment, causing nuclei to emit signals. These signals, measured after a delay, are converted into grayscale images using Fourier transformation. Adjusting RF pulses creates different MRI image types. Key parameters include Repetition Time (TR), the interval between pulse sequences, and Time to Echo (TE), the delay from RF pulse to signal reception. Moreover, tissues exhibit two important relaxation times in MRI: T1 (longitudinal) and T2 (transverse). T1 measures how quickly protons realign with the magnetic field, while T2 measures how quickly they lose phase coherence. These parameters are critical for interpreting MRI images and understanding tissue characteristics [10].

The most common MRI sequences include T1-weighted and T2-weighted scans. T1-weighted images use short TE and TR times to primarily highlight tissue contrast and brightness based on T1 properties. Conversely, T2-weighted images employ longer TE and TR times to predominantly emphasize tissue contrast and brightness based on T2 properties [10]. These sequences are distinguished by their visualization of cerebrospinal fluid (CSF), which appears dark on T1-weighted images and bright on T2-weighted images, as illustrated in Figure 1.

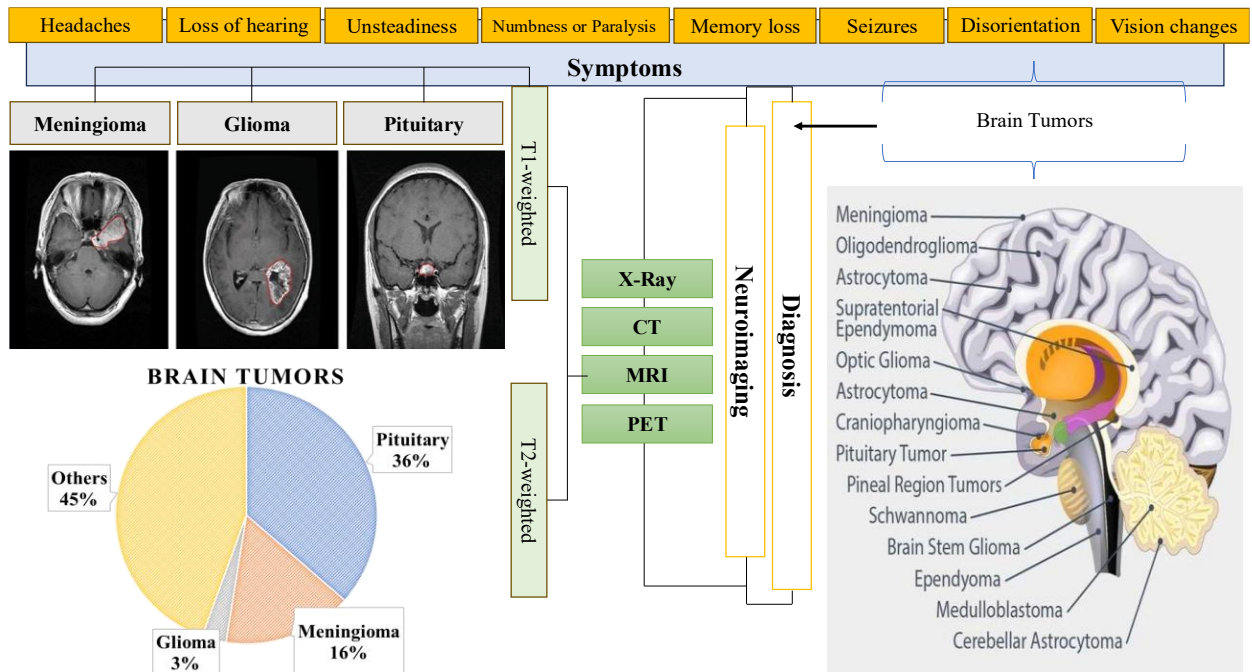

**Figure 1.** Distribution of brain tumors in AYAs: non-malignant pituitary tumors (36.5%), lower-grade gliomas (3%), and meninges tumors (16%). Includes T1 and T2 MRI images for diagnosis.

Diagnosing brain tumors promptly is critical for optimizing treatment outcomes. In the realm of IoT healthcare, researchers and medical professionals have developed several non-invasive methods to classify brain tumors and detect brain cancer. Unlike traditional detection approaches [11], computer-aided diagnostic systems (CADS) utilize Machine Learning (ML) and Deep Learning (DL) models [12], particularly Convolutional Neural Networks (CNNs) [13, 14], which are renowned for their ability to analyze medical image data with high accuracy [15]. CNNs excel in extracting pertinent features from images, thereby enhancing their utility in brain tumor classification [16]. Moreover, techniques like data augmentation and transfer learning further enhance the predictive capabilities of DL models, contributing to more precise brain tumor classification and cancer diagnosis within IoT healthcare applications [17].

Recent studies on brain tumors are systematically reviewed using databases such as the Web of Science (WoS), and their findings are comprehensively discussed in this section. One study applied a support vector machine (SVM) to 3064 MRI images from 233 patients across various tumor classes, achieving up to 94% accuracy post-preprocessing [18]. Santhosh et al. employed SVM with thresholding and watershed segmentation, achieving 85.32% accuracy in distinguishing normal and abnormal brain tissue [19]. Rafael et al. achieved an 89.6% accuracy rate using SVM for brain tumor classification [20], while Gupta and Sasidhar reported 87% accuracy using similar methods [21]. In another study, SVM achieved 92% accuracy on a dataset of 90 normal and 154 tumor brain images using advanced feature selection techniques [22]. Additionally, the RELM technique demonstrated 92.61% accuracy through rigorous cross-validation. Studies employing SVM and KNN on 102 brain tumor patients achieved accuracies of 85% and 88%, respectively [23]. These studies highlight the effectiveness of SVM and machine learning in accurately classifying brain tumors based on MRI data.

In recent years, the landscape of brain tumor diagnosis has shifted significantly from conventional machine learning methods like Support Vector Machines (SVM) to more advanced techniques such as Convolutional Neural Networks (CNNs). Historically, SVMs were prominent for their ability to classify brain tumor images based on carefully engineered features extracted from MRI scans. However, the limitations of SVMs in handling complex image features and achieving higher accuracies prompted a transition towards CNNs. CNNs have shown remarkable capability in automatically learning intricate patterns and features directly from raw images, without the need for explicit feature engineering. Studies by Choudhury et al. [24], Martini and Oermann [25], and Afshar et al. [26] have highlighted this shift. They achieved accuracies ranging from 90.89% to 96.08% using CNNs for brain tumor classification tasks, showcasing the effectiveness of deep learning in enhancing diagnostic accuracy.

In recent advancements for brain cancer detection, two prominent CNN architectures, U-Net and Fast R-CNN, have emerged as pivotal tools. U-Net excels in semantic segmentation tasks by efficiently capturing intricate details in medical images, particularly for delineating tumor boundaries. On the other hand, Fast R-CNN integrates region-based convolutional networks to accurately classify tumors by identifying relevant regions of interest within MRI scans. These CNN models represent significant strides in enhancing the accuracy and efficiency of brain cancer detection through advanced image processing techniques.

Le and Pham [27] proposed utilizing full-convolution U-Net networks to extract features from multimodal MRI training datasets, subsequently employing Extremely Randomized Trees (ExtraTrees) for abnormal tumor cell segmentation. Wang et al. [28] implemented a triple-cascaded architecture using three binary classifiers, enhancing robustness and simplifying training, albeit with longer training times due to its non-end-to-end nature. Dong et al. [29] developed a deep convolutional network based on U-Net for automatic tumor segmentation, evaluated on BRATS' 2015 dataset, utilizing the Dice loss function. In 2018, Chen introduced a separable 3D U-Net architecture that enhances computational efficiency by parallelizing each 3D convolution into three branches [30].

Recent studies have demonstrated significant advancements in brain tumor detection using Region-based Convolutional Neural Networks (R-CNNs) and their variants. Kumar et al. [31] integrated Residual Networks (ResNet-101) with Faster R-CNN to enhance accuracy and efficiency in tumor detection with 95% accuracy, leveraging the network's ability to capture complex patterns from diverse brain images. Rao et al. [32] utilized Faster R-CNN to classify intracranial tumors like meningioma and glioma, highlighting the technique's computational speed and accuracy in differentiating between tumor types in MRI images with 97.86% accuracy. Ezhilarasi et al. [33] combined AlexNet with Region Proposal Network (RPN) in a Faster R-CNN framework, employing transfer learning to enhance the classification accuracy of different tumor types.

Transitioning from U-Net and R-CNN, it is crucial to highlight the integration of transfer learning, which significantly enhances the performance of these methods in medical image analysis. Transfer learning leverages pre-trained models on large datasets, enabling the creation of high-performing models with reduced training time and computational cost [34]. This technique is especially valuable in the medical field, where annotated datasets can be limited. Srinivas et al. [35] conducted a comprehensive study comparing the performance of transfer learning-based CNN models pre-trained with VGG16, ResNet-50, and InceptionV3 architectures for brain tumor cell prediction. Their results demonstrated that VGG16 achieved a high accuracy of 96%, ResNet-50 reached 95%, and InceptionV3 had an accuracy of 78%. Additionally, Khan et al. [36] proposed a brain tumor classification method employing transfer learning techniques, achieving an accuracy of 94.82%. These studies underscore the effectiveness of transfer learning in enhancing the accuracy and efficiency of brain tumor classification, making it a crucial tool in the advancement of medical diagnostics.

In our study, we compared R-CNN, U-Net architecture, and Fully Connected CNN with three transfer learning models—Inception-V3, EfficientNetB4, and VGG19—to determine the most effective approach for classifying brain MRI data. Performance metrics such as Accuracy, Precision, Recall, and F-score are calculated based on the confusion matrix to evaluate the performance of classification models used in tasks like brain tumor detection in medical image processing. Our primary contribution lies in the innovative application of transfer learning and fine-tuning of MRI data to categorize brain tumors into three groups. Our key contributions include:

- **Data Preprocessing and Feature Extraction**: Unlike most recent works, we preprocess the test dataset by extracting features regionally and pixel-wise before applying the model's weights and biases for overall evaluation.
- **Enhanced R-CNN Performance**: By using selective search methodology in Fast R-CNN, we fine-tuned the transfer learning models, resulting in higher precision in feature extraction and efficient performance in both tumor classification and non-labeled datasets.
- **Optimized U-Net Architecture**: We fine-tuned the U-Net model for the three classifications through down-sampling and up-sampling as feature extraction methodologies, analyzing their efficacy in model performance.
- **Fine-Tuning Transfer Learning Models**: We processed and fine-tuned transfer learning models, applying them to three benchmark datasets. Enhancements to models like Inception-V3, EfficientNetB4, and VGG19 included adding a single fully connected layer to optimize performance.
- **Benchmark Comparison**: We established a benchmark to evaluate our proposed transfer learning methodologies against previous research, achieving maximum precision with VGG19-based CNN at 97.5% and InceptionV3 at 98.8% in operational scenarios.

- **Leveraging Pre-Trained Models**: Transfer learning allows us to leverage pre-trained models, which is especially beneficial when dealing with limited labeled medical data.
- **Impact on Diagnostic Accuracy**: Our results highlight the effectiveness of our transfer learning methodologies in classifying brain tumors, demonstrating their potential impact on advancing diagnostic accuracy in medical image analysis.

This article is structured as follows. Following the review of AI applications in medical diagnostics presented in this introduction, the Materials and Methods section explains the proposed methodology in detail and describes the experimental setup, covering data preparation, model training, and evaluation processes. In the Results and Discussion section, we present the performance outcomes of the model and provide a comprehensive analysis of the findings. Finally, the Conclusion and Future Work section summarizes the key insights and proposes potential directions for future investigations.

**Method**

This section provides an overview of the dataset and methodologies employed in the study. The dataset consists of three types of brain tumors: meningioma (708 slices), glioma (1426 slices), and pituitary tumor (930 slices). Classification tasks were tackled using Fast Region-based Convolutional Neural Network (R-CNN) and U-Net architectures. Furthermore, transfer learning models such as VGG19, VGG16, InceptionV3, and EfficientNetB4 were adapted and integrated to enhance classification accuracy based on these frameworks.

The dataset comprises a total of 3064 T1-weighted contrast-enhanced MRI images of the human brain, systematically classified into three distinct categories: class 0 for meningioma, class 1 for glioma, and class 2 for pituitary tumor. The distribution of labeled images across these three classes is detailed in *Table 1* [37]. These classes are labeled by physicians. The size of the input images is 270 × 270. Table 1 shows the training, test, and validation set discriminations by class. The flow diagram of our models is illustrated in *Figure 2*, underlining the architectures used to improve the accuracy and efficiency of brain tumor classification for better diagnostic outcomes in the medical field.

| Data | Meningioma | Glioma | Pituitary | Total |
|---|---|---|---|---|
| Class type | 0 | 1 | 2 | 3 classes |
| Training data | 437 | 936 | 588 | 1961 |
| Cross-validation data (stratified k-Fold (k=5)) | 109 | 234 | 147 | 490 |
| Test data | 162 | 256 | 195 | 613 |

**Table 1.** Overview of the Preprocessed Brain Tumor Dataset Distribution.

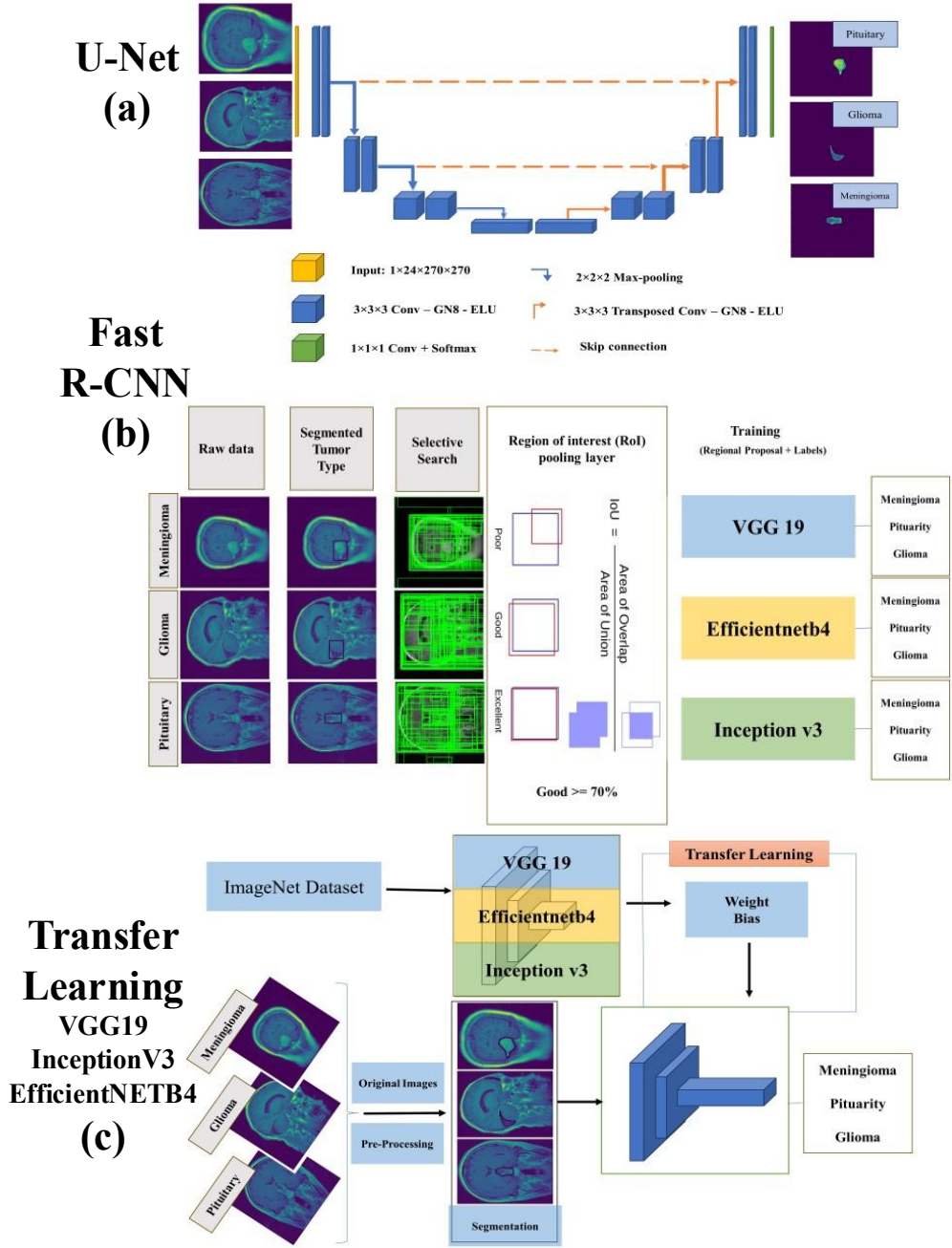

**Figure 2.** Flow diagram depicting **(a)** Fast R-CNN, **(b)** U-Net, and **(c)** Transfer Learning-based CNN architectures.

Convolutional neural networks (CNNs) represent a pivotal advancement in deep learning and computer vision. These architectures are tailored to extract intricate features from complex visual data, such as images and video. The CNN's structure, comprising convolutional layers, pooling layers, and fully connected layers, mirrors the human visual system's capacity to discern patterns and hierarchical features. Convolutional layers employ convolution operations to detect local features, which are subsequently abstracted by pooling layers to distill essential information. These hierarchical representations are then processed by fully connected layers for tasks like classification or regression [38]. CNNs have revolutionized image recognition, demonstrating significant achievements across various domains including image classification, object detection, face recognition, and medical image analysis [1].

UNets derive their name from their architectural resemblance to the letter "U," comprising both an encoder and a decoder [39]. The encoder extracts high-level features from input data, while the decoder maps these features to specific outputs. Residual links within UNets prevent gradient vanishing issues commonly encountered in segmentation tasks, ensuring robust performance [40]. When UNets incorporate branches connected to other networks for diverse tasks, they become hybrid architectures capable of multitasking [41]. In our proposed network, the encoder is integrated with a fully connected network for classification, making it a hybrid architecture designed for multiple objectives.

Fast Region-based Convolutional Neural Network (Fast R-CNN) is an advanced object detection algorithm that builds upon the original R-CNN (Region-based Convolutional Neural Network), enhancing both speed and accuracy in detecting objects within images. The conventional R-CNN method processed each region proposal independently using a CNN, resulting in high computational costs and slower processing times. Fast R-CNN introduces several innovations to overcome these challenges. It incorporates a Region Proposal Network (RPN) that generates region proposals directly from the convolutional feature map. The RPN operates by sliding a small network over the feature map, predicting scores and bounding box coordinates for potential regions. This approach significantly streamlines the detection process by eliminating the need for external region proposal methods like selective search, thus improving efficiency without compromising accuracy. In our study, depicted in **Figure 2**, Fast R-CNN with RPN was employed across various image classes, underscoring its role in swiftly identifying and localizing objects of interest.

Intersection over Union (IoU) is a critical evaluation metric employed in Fast R-CNN to assess the accuracy of detected bounding boxes. It quantifies the overlap between a predicted bounding box and the ground truth bounding box by calculating the ratio of their intersection area to their union area. A higher IoU value signifies a more precise prediction. IoU plays a pivotal role in distinguishing between true positive and false positive detections and is often utilized with predefined thresholds to determine if a predicted bounding box sufficiently matches the ground truth. In our study, illustrated in **Figure 2**, we considered IoU values greater than 0.7 as indicative of reliable detection candidates. Fast R-CNN, renowned for its prowess in object detection and classification within images, contrasts with U-Net models commonly utilized in medical imaging for pixel-wise segmentation tasks. While Fast R-CNN excels in identifying and categorizing regions of interest, U-Net models specialize in the precise delineation of object boundaries, particularly beneficial for tasks like tumor segmentation in medical images.

In our study, we used the transfer learning approach, which is a cornerstone in both machine learning and deep learning, leveraging knowledge gained from training a model on one task to enhance performance on another related task. In neural networks, transfer learning is particularly powerful, involving the use of a pre-trained model initially trained on a diverse dataset. This model is then fine-tuned on a new dataset or task to improve its accuracy and effectiveness [42]. In this study, the classification process utilized transfer learning models including VGG19, InceptionV3, and EfficientNetB4.

VGG19 represents an advanced convolutional neural network architecture, expanding upon the original VGG16 model with 19 layers including 16 convolutional and 3 fully connected layers. It utilizes 3 × 3 convolutional filters to effectively capture intricate patterns and features from image data. Employing max-pooling layers, VGG19 reduces spatial dimensions, thereby enhancing computational efficiency. Its final fully connected layers enable high-level feature extraction for accurate predictions. Activation functions like Rectified Linear Units (ReLU) introduce nonlinearity, contributing to its effectiveness in image classification tasks. Despite its historical significance, VGG19 has been surpassed by more recent architectures such as ResNet and Inception in terms of performance and computational efficiency [43].

InceptionV3, developed by Google, is an advanced convolutional neural network architecture specifically designed for efficient image classification tasks. It represents an evolution from its predecessors, InceptionV1 (GoogLeNet) and InceptionV2, incorporating innovative features such as inception modules [1]. These modules utilize parallel convolutions, pooling layers, and other operations to capture diverse spatial features concurrently. InceptionV3 also integrates factorized convolutions, auxiliary classifiers, batch normalization, the RMSProp optimizer, and label smoothing techniques to enhance both accuracy and computational efficiency [43]. Compared to VGG19, which relies on deep networks of 3x3 convolutions and is computationally intensive, InceptionV3 achieves superior performance with fewer parameters, enabling faster training and making it well-suited for large-scale image classification tasks [42].

EfficientNetB4, developed by Google AI, is a state-of-the-art convolutional neural network renowned for its exceptional balance between accuracy and computational efficiency in image classification tasks. It employs a sophisticated approach known as compound scaling to uniformly adjust the network's depth, width, and resolution, optimizing performance without imposing significant computational overhead. Key features include the Swish activation function, enhancing gradient flow and optimization, depth-wise separable convolutions to minimize parameters and computational costs, and squeeze-and-excitation (SE) blocks for refining feature focus [44].

Compared to VGG19, which utilizes traditional 3x3 convolution layers and requires higher computational resources, EfficientNetB4 achieves comparable or superior accuracy with enhanced computational efficiency. In contrast to InceptionV3, noted for its multi-scale feature extraction using inception modules and factorized convolutions, EfficientNetB4's design prioritizes computational efficiency through compound scaling and depth-wise separable convolutions. This positions EfficientNetB4 as ideal for applications demanding high accuracy under limited computational resources, while InceptionV3 excels in robust feature extraction and comprehensive feature representation.

**Results**

This study explores the challenges of image classification using deep learning (DL) techniques, with a particular focus on medical images, specifically brain tumor classification. Five models through five tasks—3-Layer CNN, U-Net, Fast R-CNN, InceptionV3, EfficientNetB4, and VGG19—were evaluated for their performance in this task. The evaluation utilized a standardized dataset split into training (80%), validation (stratified cross-validation with k=5), and testing (20%) subsets. All experiments and analyses were conducted on the Kaggle platform, leveraging a dual GPU setup featuring P100 GPUs.

As the initial task of this study, we employed a 3-layer CNN, as shown in **Figure 3**, architecture specifically designed for the classification of brain tumor images resized to 270 × 270 pixels. The architecture begins with a foundational 3 × 3 convolutional layer featuring 16 filters, followed by 2 × 2 max pooling to facilitate down sampling. Sequential layers incorporate 32 and 64 filters respectively, with integrated L2 regularization to enhance model generalization and mitigate overfitting. Each convolutional layer is accompanied by max pooling, strategically applied to emphasize critical features within the data. To further stabilize training and promote generalization, a dropout layer with a rate of 0.5 is introduced after the initial pooling layer, complemented by batch normalization.

The resulting flattened feature maps are propagated through dense layers comprising 16 units each, integrated with batch normalization to maintain stable training dynamics. The architecture culminates in a final dense layer equipped with 3 units and softmax activation, facilitating precise classification of brain tumor types.

Training of the model employs sparse categorical cross-entropy loss optimized through the Adam optimizer with a learning rate set at 0.001. The dataset is partitioned into training (80%) and testing (20%) sets, with 5-fold stratified cross-validation employed to validate the model's robustness and generalizability across diverse subsets. All computational experiments are conducted on the Kaggle platform, leveraging GPU P100 resources to ensure efficient execution and thorough performance evaluation.

**U-Net Architecture**

The U-Net architecture, depicted in **Figure 2** (a), stands as a sophisticated convolutional neural network widely acclaimed for its efficacy in image segmentation, particularly in the realm of medical imaging such as brain tumor classification. This specialized U-Net model is meticulously crafted to classify grayscale images into three distinct classes, each standardized to 270x270 pixels for consistency. Beginning with an encoder pathway, the model integrates multiple Conv2D layers, progressively increasing in filter count (64, 128, 256, 512), to meticulously extract hierarchical features from the input images. Each Conv2D layer is complemented by a Rectified Linear Unit (ReLU) activation function, maintaining spatial integrity with 'same' padding.

MaxPooling2D layers follow each pair of Conv2D layers, strategically downsampling feature maps (2x2 pooling size) to capture essential details while reducing spatial dimensions. This process ensures that critical information is preserved for subsequent analysis. At the network's bottleneck (conv5), the feature maps are further refined through 1024 filters, enhancing the model's discriminative ability across different classes. Additional Conv2D layers with ReLU activation and 'same' padding maintain spatial consistency throughout the network.

GlobalAveragePooling2D compresses the bottleneck output into a flat vector, preparing the feature maps for classification by effectively summarizing the extracted features. Dropout regularization (Dropout(0.5)) is incorporated to prevent overfitting during training, randomly deactivating a portion of input units to enhance model generalization.

The final layer of the model, a Dense layer (Dense (num_classes, activation='softmax')), employs softmax activation to compute class probabilities based on the learned features, enabling robust predictions for the three classes. During training, the Rectified Adam optimization algorithm (Adam(learning_rate=0.001)) dynamically adjusts the learning rate to optimize gradient flow and enhance training efficiency. Categorical cross-entropy serves as the loss function, ideal for multiclass classification tasks, with metrics such as accuracy, precision, and recall monitored to assess model performance.

To ensure stable training, a learning rate reduction strategy (ReduceLROnPlateau) is implemented, adapting the learning rate as necessary based on the loss function's progression during training. The model undergoes 14 epochs of training with a batch size of 10, facilitating efficient learning and convergence while showcasing the network's capability in extracting intricate features crucial for accurate brain tumor classification.

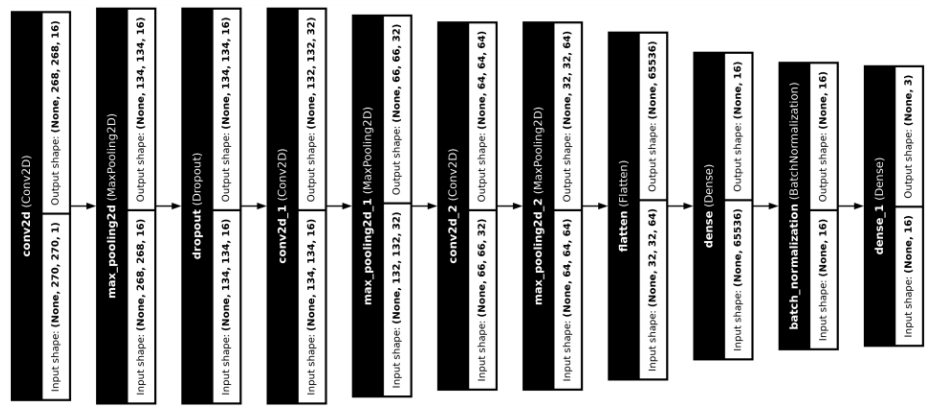

**Figure 3**. 3-Layer CNN Architecture.

**Fast R-CNN Architecture**

Fast R-CNN has emerged as a significant advancement in object detection models, offering enhanced efficiency and accuracy, particularly in applications like brain cancer segmentation. Unlike traditional R-CNN, which involves a multi-stage pipeline and slower training times, Fast R-CNN optimizes the process by integrating region of interest (RoI) pooling and performing classification directly on these regions within a single network pass. This reduces computational overhead and training time, making it well-suited for large datasets such as those encountered in medical imaging.

For brain cancer segmentation, as *Figure 2(b)* after preprocessing and employing effective segmentation methodologies, we applied selective search to identify a comprehensive set of candidate regions. These regions were filtered using Intersection over Union (IoU) with a threshold of 0.7, ensuring the selected regions are closely aligned with ground truth. Fast R-CNN's RoI-based approach allows the model to assign varying weights to different regions of the image, focusing more on areas critical for identifying tumor boundaries. This is particularly beneficial for non-convex problems like cancer classification, where precise boundary delineation is essential. The results demonstrate that Fast R-CNN outperforms traditional methods in terms of both accuracy and speed, making it a robust solution for complex medical image segmentation tasks.

**CNN-based transfer learning**

Our model utilizes transfer learning with InceptionV3 pretrained on ImageNet for classifying brain tumors into three categories, as shown in **Figure 4**. We initially froze early layers to retain learned features, followed by a Flatten layer, BatchNormalization, and Dropout to enhance generalization. A Dense layer with softmax activation predicts tumor classes. Training used sparse categorical cross-entropy loss, Adam optimizer (lr=0.001), and class weights for balanced representation. Stratified K-Fold cross-validation with 5 folds ensured robust evaluation across varied data splits, monitoring metrics like loss and accuracy over 30 epochs per fold. This approach leverages transfer learning effectively, adapting InceptionV3's features for precise brain tumor classification and offering insights into model performance and potential clinical use.

In this study, we employed transfer learning using EfficientNetB4 pretrained on ImageNet for the task of classifying brain tumor images into three distinct categories as shown in **Figure 4**. The initial layers of EfficientNetB4 were frozen to retain the learned features from the large-scale image dataset. We appended a Flatten layer to transform the output of the base model into a vector format, followed by BatchNormalization for standardization and Dropout regularization (20%) to mitigate overfitting. A Dense layer with softmax activation was added to produce class predictions. The model was compiled using sparse categorical cross-entropy as the loss function and Adam optimizer with a learning rate of 0.001. We tracked model performance metrics, including accuracy, throughout training. Given the class imbalance in our dataset, we computed class weights to provide higher influence to underrepresented classes during training, enhancing model robustness. Moreover, stratified K-Fold cross-validation with 4 folds was employed to evaluate the model's performance across different data partitions, ensuring reliable generalization. Each fold was trained for 30 epochs with a batch size of 32, utilizing both training and validation datasets for model assessment and optimization.

**VGG19**

This study utilizes the VGG19 architecture pre-trained on the ImageNet dataset to classify brain tumor images into three distinct categories. VGG19 is renowned for its deep convolutional neural network structure comprising 16 convolutional layers and 3 fully connected layers, designed to extract intricate features from complex image data as shown in **Figure 4**. The initial layers of the pre-trained VGG19 model were frozen to leverage the learned representations from ImageNet, thereby enhancing feature extraction capabilities.

Following the convolutional layers, a flattened layer was added to convert the output into a vector format, facilitating subsequent processing. BatchNormalization was applied to standardize the feature maps, and Dropout regularization with a rate of 20% was implemented to mitigate overfitting during training. The model concludes with a Dense layer employing softmax activation, which outputs probabilities for each of the three tumor categories. For training, the model was compiled using sparse categorical cross-entropy as the loss function and Adam optimizer with a fixed learning rate of 0.001. To address class imbalance within the dataset, class weights were computed to ensure equitable learning across all categories during training.

To assess the model's performance robustness and generalizability, Stratified K-Fold cross-validation with 4 folds was employed. This validation technique partitions the dataset into subsets while maintaining class distribution integrity, ensuring unbiased evaluation across diverse data segments. Each fold underwent 30 epochs of training, using a batch size of 32, and incorporated both training and validation datasets for iterative model assessment and optimization.

**Figure 5** and **Figure 6** display the training and validation accuracy loss graphs for models developed using 3-Layer CNN, U-Net, Fast R-CNN, InceptionV3, EfficientNetB4, and VGG19 architectures.

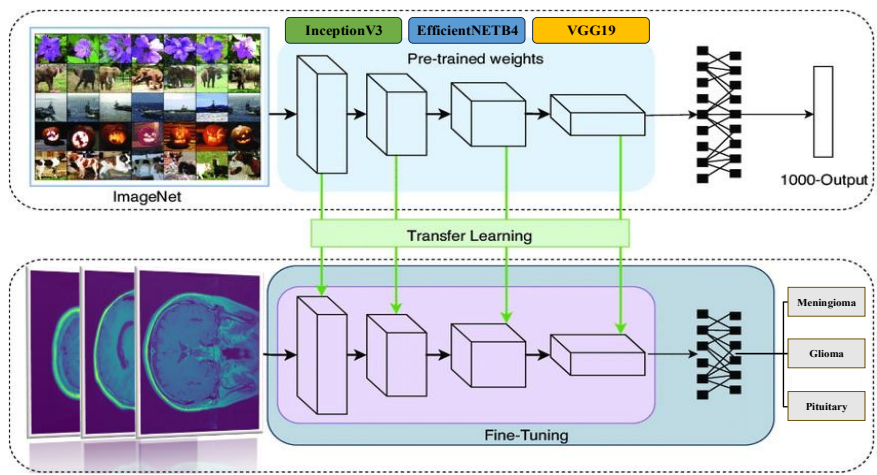

**Figure 4.** Transfer Learning base-CNN (EfficientNETB4, VGG19-CNN, InceptionV3) Architectures.

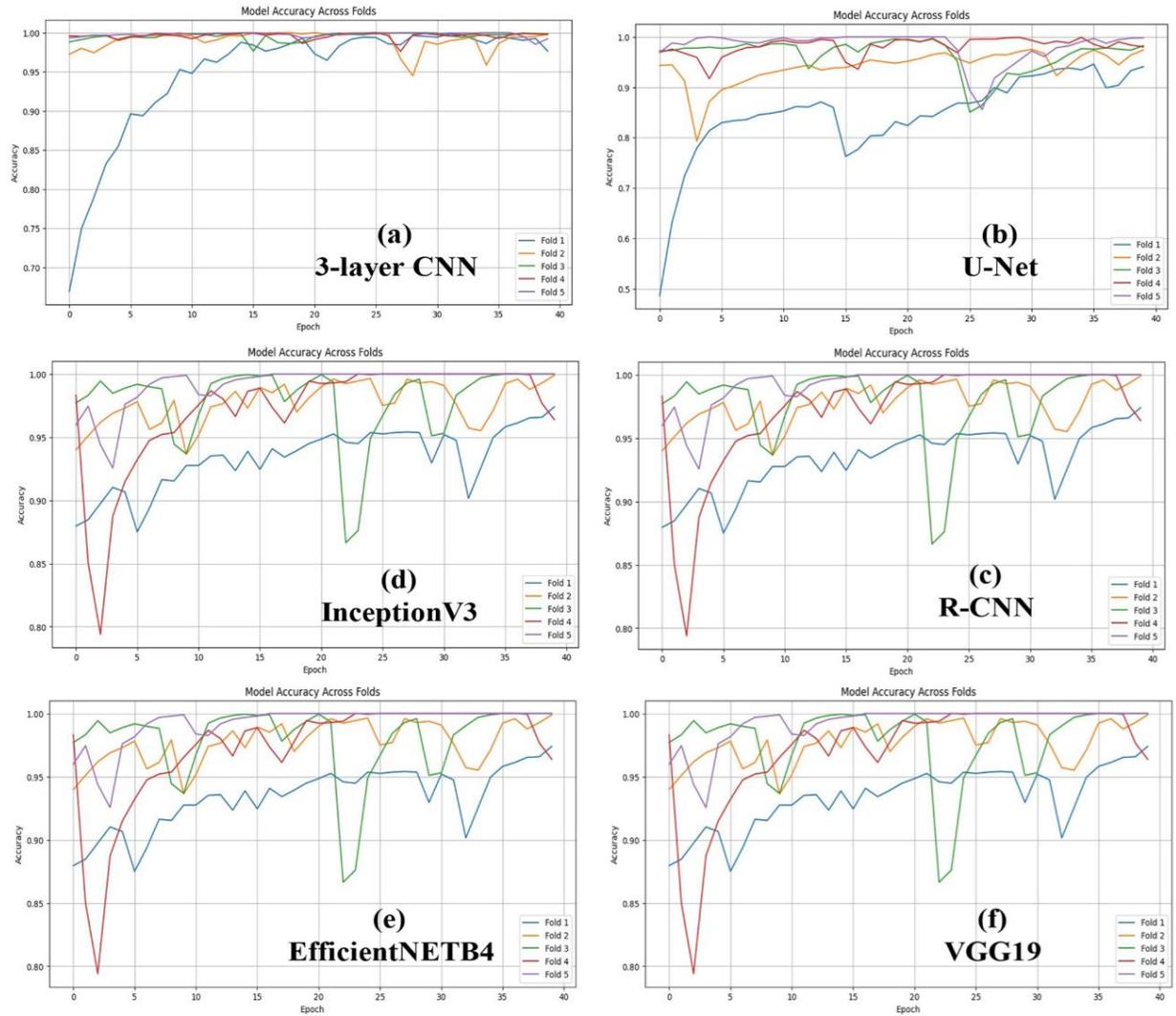

**Figure 5.** Accuracy plots of our proposed models, (a) 3-layer CNN, (b) U-Net architecture, (c) TL-EfficientNETB4-CNN, (d) TL-VGG19-CNN, and (e) TL-InceptionV3-CNN.

| Models | Accuracy | F-score | Recall | Precision | AUC |
|---|---|---|---|---|---|
| 3-Layer CNN | 99 | 99 | 99 | 99 | 99.33 |
| U-Net | 99.56 | 99 | 99 | 99 | 99.8 |
| InceptionV3 | 97.27 | 97 | 97 | 97 | 99 |
| EfficientNETB4 | 98.31 | 98 | 98 | 98 | 99.33 |
| VGG19 | 97.72 | 97 | 97 | 97 | 99 |

**Table 2**. Task Models' Performance.

The study also presents the accuracy, F-score, recall, precision, and AUC results obtained from these models in the study. According to classification. The ROC curves for the models, as depicted in **Figure 7**, illustrate distinct advantages for the 3-layer CNN, U-Net, and the transfer learning models VGG19, InceptionV3, and EfficientNetB4. The AUC values highlight the robust classification capabilities of these models.

*Table* **2**, U-Net achieved the highest accuracy at 99.5%, surpassing other models. Its F-score was 98%, AUC value 99.8%, recall 98%, and precision 98%. These metrics underscore U-Net's superior performance in brain tumor

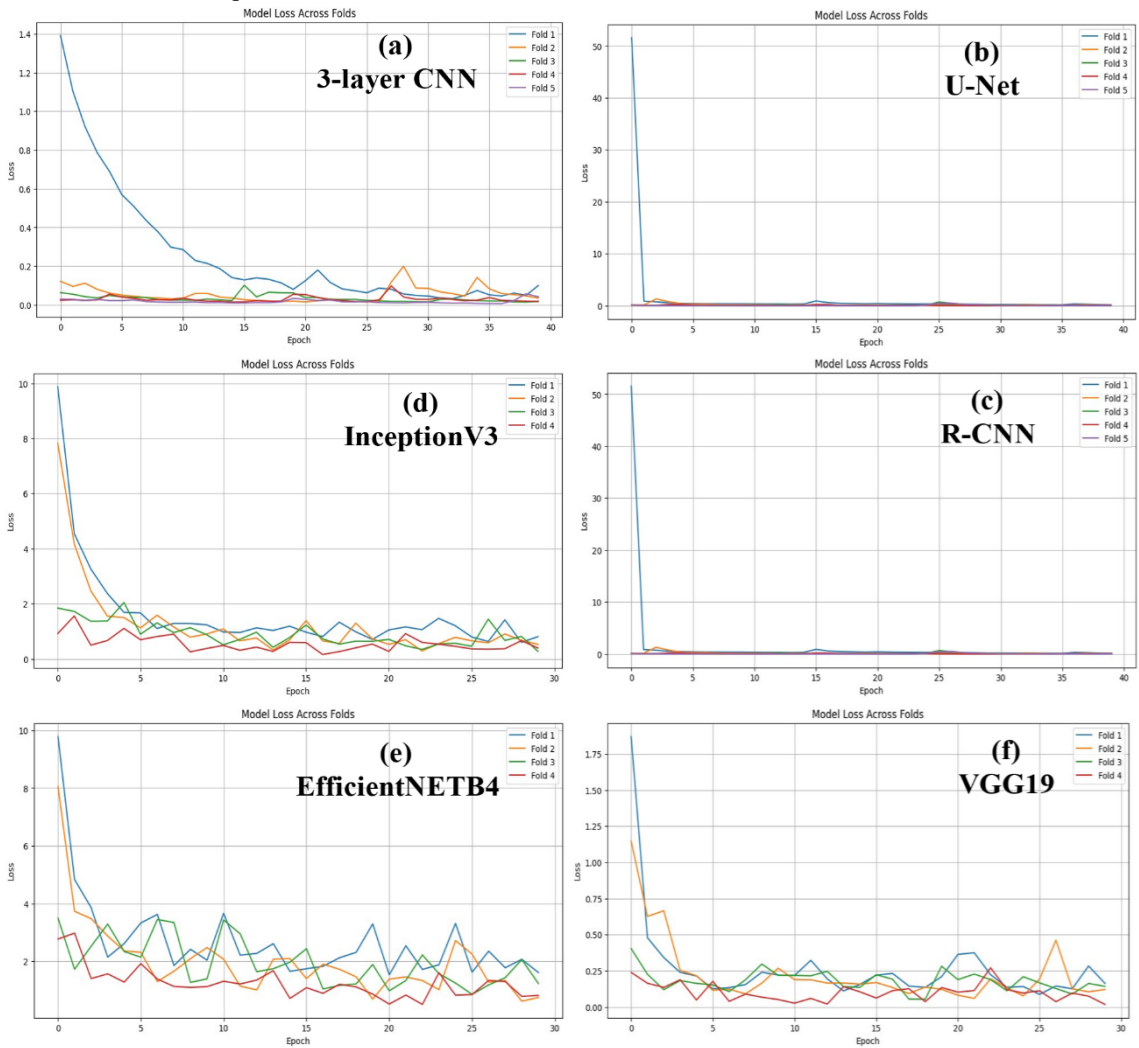

**Figure 6.** Loss plots of our proposed models, (a) 3-layer CNN, (b) U-Net architecture, (c) TL-EfficientNETB4-CNN, (d) TL-VGG19-CNN, and (e) TL-InceptionV3-CNN.

*Figure 8* presents the confusion matrix for classifying glioma, meningioma, non-tumor normal patients, and pituitary tumor patients. The high classification performance across all models—3-layer CNN, U-Net, and transfer learning models VGG19, InceptionV3, and EfficientNetB4—demonstrates the efficacy of our approach.

These results validate our integrated framework, emphasizing the potential of advanced deep-learning techniques in medical image analysis. The utilization of U-Net and transfer learning-based CNN models not only enhances classification accuracy but also ensures reliability in distinguishing between different types of brain tumors. This study provides a strong foundation for further research and clinical application, highlighting the importance of leveraging sophisticated neural network architectures for improved diagnostic outcomes in healthcare.

While AI technology is not intended to replace doctors, it serves to support and enhance their work significantly. To fully realize the potential of AI, it is crucial to address issues such as ethics, security, and privacy. AI-based solutions will continue to contribute to the effective management of brain tumors and other health conditions, ultimately improving patients' quality of life. As demonstrated in this study, AI will increasingly play a vital role in human health, from early diagnosis to facilitating positive treatment outcomes. Our findings suggest that transfer learning methods are particularly effective for image processing-based applications in healthcare, aiding decision-makers with timely and accurate diagnoses. The data derived from MRI or CT scans can act as an early warning system, providing health professionals with the tools they need for swift and precise decision-making. Thus, alongside empirical analysis, AI-based applications should assume a more prominent role promptly. The efficacy of AI models, especially the U-Net architecture, in brain tumor classification is noteworthy. U-Net achieved the highest accuracy at 99.5%, with an F-score of 98%, an AUC value of 99.8%, recall at 98%, and precision at 98%. These metrics underscore the robustness of U-Net in distinguishing between different types of brain tumors, making it a valuable tool in medical diagnostics. In the future, we will explore the diagnosis of diseases from real-time CT or MR images. By leveraging the advanced capabilities of models like U-Net and transfer learning-based CNN architectures, we aim to enhance the accuracy and reliability of early diagnosis systems. This approach not only facilitates better patient outcomes but also supports healthcare professionals in delivering more effective and timely interventions. The continued development and integration of AI in medical diagnostics promise significant advancements in patient care and treatment efficacy.

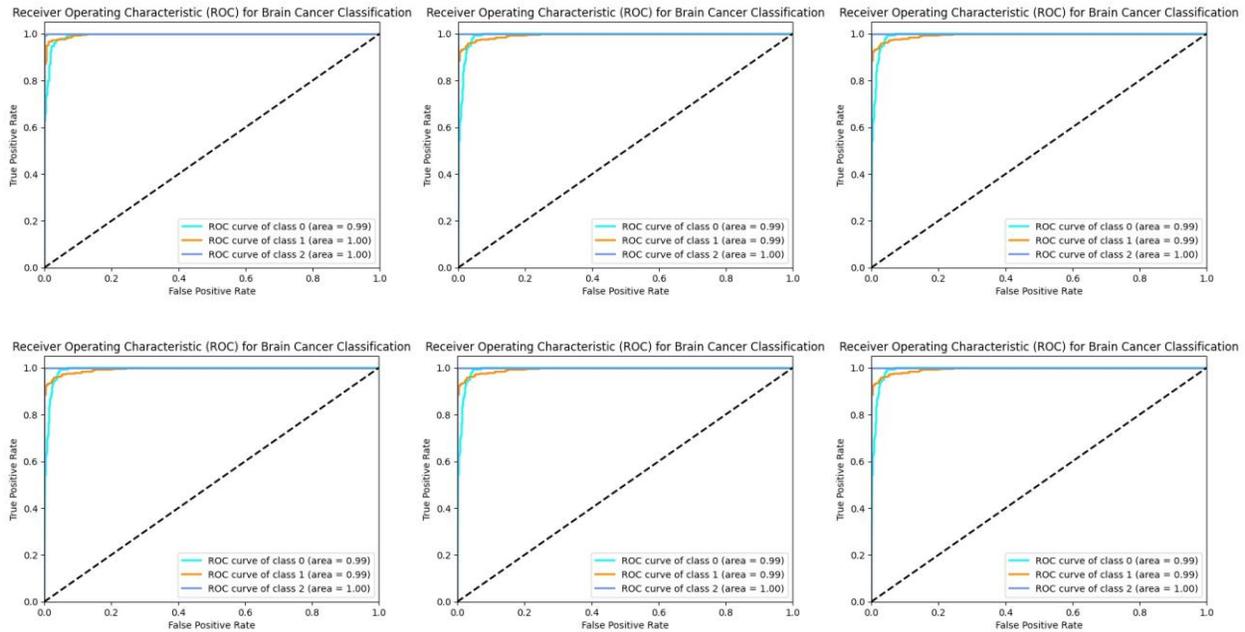

**Figure 7.** Receiver Operating Characteristic - Area Under the Curve.

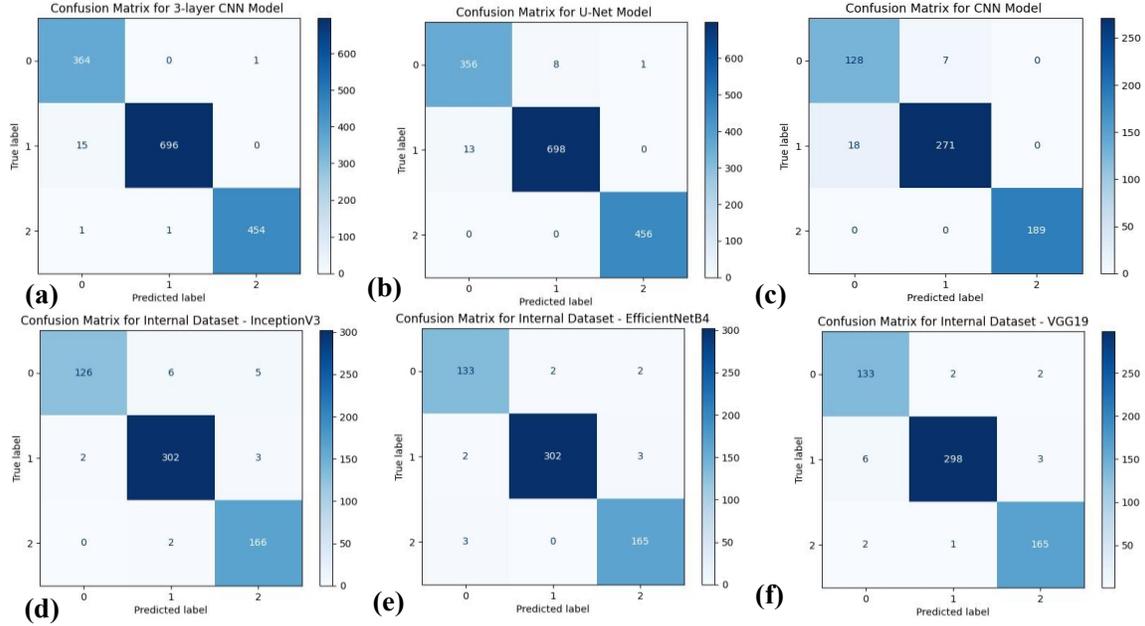

**Figure 8.** Confusion Matrix

| Model | Methodology | Accuracy | Ref. |
|---|---|---|---|
| SVM | combination of thresholding and watershed segmentation techniques | 85.32% | [19] |
| | Feature extraction using a regularized extreme learning machine (RELM) | 92% | [22] |
| | | | [23] |
| KNN | - | 88% | |
| CNN | - | 96.08% | [24] |
| | - | 93.9% | [25] |
| | using GIST, Normalized GIST (NGIST), and PCA-NGIST methods | 91.43% | [23] |
| | Capsule Network (CapsNet) | 90.89% | [45] |
| | | 90.89% | [46] |
| | developing an IoT computational system based on deep learning | 89% | [47] |
| LSTM | | 90.02% | |
| CNN-LSTM | | 92% | |
| Transfer learning-based CNN | VGG16 | 96% | [35,36] |
| | pre-trained CNN | 94.82% | |
| | ResNet-50 | 95% | |
| | InceptionV3 | 78% | |
| | Hyperparameters of the Residual Network 50 (ResNet50) model (I-HGS-ResNet50) | 99.89% | [47] |
| GA-CNN | genetic algorithm (GA)+CNN | 94.2% | [48] |
| U-Net | Cascaded Anisotropic Convolutional Neural Networks | 0.7859%, 0.9050%, 0.8378% | [28] |
| | Extremely randomized trees (ExtraTrees) classifier for segmentation | 72-85% | [27] |
| R-CNN | Integrating Residual Networks (ResNet-101) with Faster R-CNN | 95% | [31] |
| | Faster R-CNN | 97.86% | [32] |

**Table 3.** Model's performance of recent works.

**External cohort cross-dataset validation**

The proposed technique underwent cross-dataset validation using an external dataset to further demonstrate its robustness. For this purpose, a brain tumor classification dataset was obtained from the open-source Kaggle repository. The proposed model was not trained on normal instances, so the normal class was excluded from the dataset for cross-validation. *Table 4* summarizes the results from this cross-dataset validation on the external cohort, along with their confusion matrix plots in *Figure 9* and ROC results in **Figure 10**. Among the various fine-tuned EfficientNet models, EfficientNetB2 achieved the highest performance, with precision, recall/sensitivity, specificity, F1-score, and accuracy values of 92.11%, 92.11%, 95.96%, 92.02%, and 92.23%, respectively. These results highlight the significant resilience of the suggested modifications to EfficientNets when applied to an external dataset.

| Models | Accuracy | F-score | Recall | Precision | AUC |
|---|---|---|---|---|---|
| 3-Layer CNN | 94.94 | 95 | 96 | 95 | 97 |
| EfficientNETB4 | 91.07 | 90 | 90 | 90 | 96.33 |
| VGG19 | 93.35 | 91 | 91 | 91 | 97.33 |
| InceptionV3 | 86.46 | 86 | 86 | 86 | 93.33 |
| U-Net | 96.01 | 96 | 96 | 96 | 99.33 |

**Table 5.** External cohort cross-dataset validation on fine-tined models.

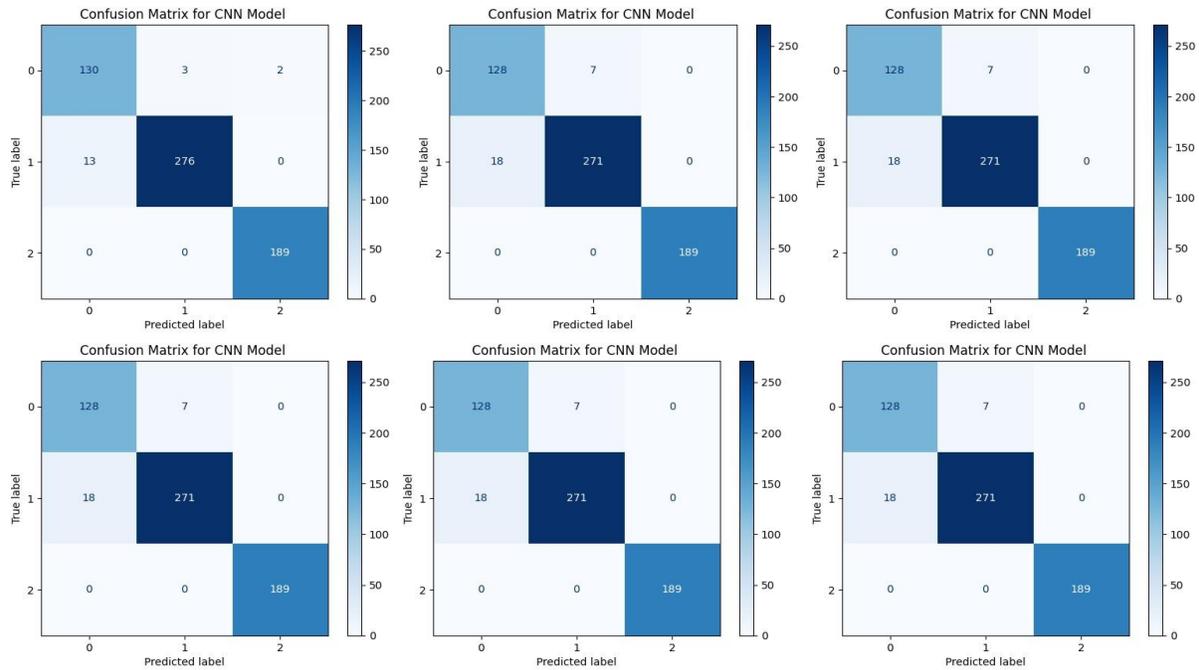

**Figure 9.** Confusion Matrix on external cohort cross-dataset validation.

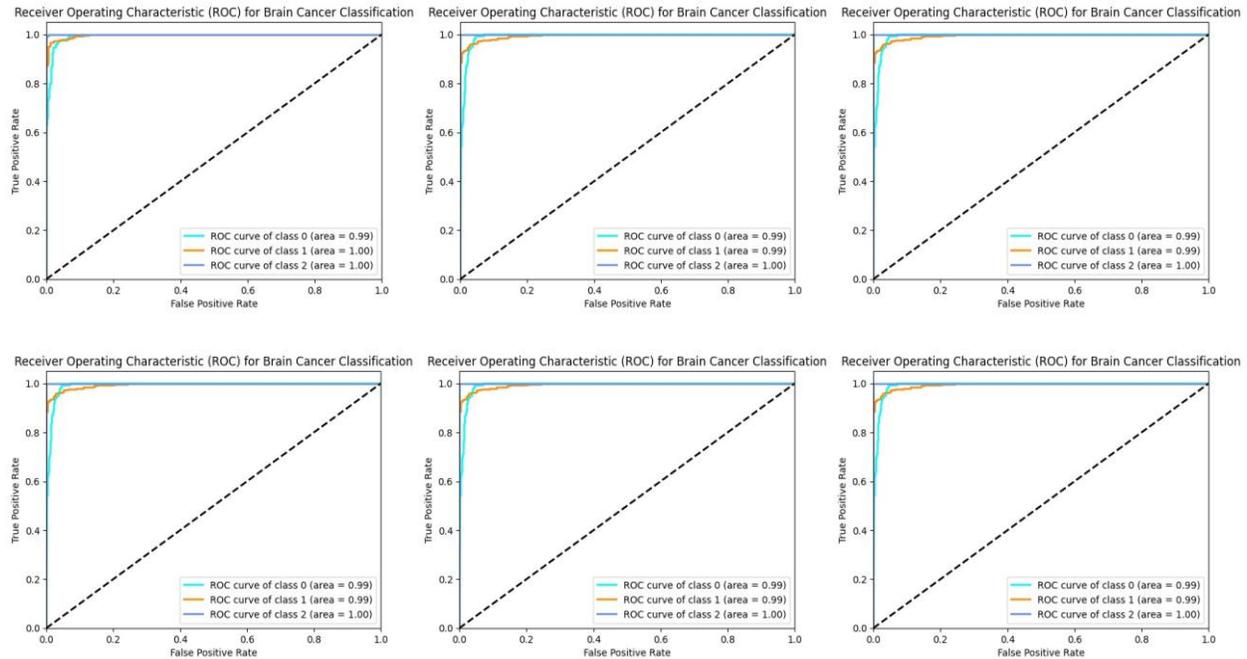

**Figure 10.** Receiver Operating Characteristic - Area Under the Curve.

**Limitation**

In our study, we applied multi-layer CNN, U-Net, and transfer learning-based CNN models (EfficientNetB4, VGG19, and InceptionV3) for brain tumor detection. While these methods show promise, several common issues arise. Firstly, the dependency on large, labeled datasets for training is a significant limitation. Medical imaging datasets are often small and hard to obtain due to privacy concerns and the complexity of annotating medical images, leading to potential overfitting and reduced generalizability of the models. Secondly, variability in imaging protocols and quality across different medical institutions can introduce inconsistencies in the data, complicating model training and evaluation. Models may perform well on specific datasets but struggle to generalize across different imaging conditions. Thirdly, these deep learning models are computationally intensive, requiring significant resources for training and fine-tuning, which may not be available in all healthcare settings, limiting their practical deployment. Lastly, the interpretability of these models remains a challenge. As "black box" systems, they provide little insight into their decision-making process, hindering clinical adoption where clear and explainable results are essential. Addressing these limitations is crucial for advancing the application of deep learning in brain tumor detection and ensuring reliable, widespread use in clinical practice. Future research should focus on improving data augmentation, model interpretability, and computational efficiency.

**Future work**

For future research, we propose investigating the use of region-based CNNs and deep segmentation techniques to improve the detection of brain tumors. This approach could potentially enhance the accuracy and specificity of tumor detection by focusing on localized regions of interest within the brain images. Moreover, there is a promising avenue to explore the application of transformer-based architectures for brain tumor classification. Transitioning from traditional deep CNN methods to transformer models could offer a novel perspective, leveraging their ability to capture long-range dependencies and global context. This shift has the potential to provide more comprehensive and information-rich feature maps, thereby improving the model's pattern recognition capabilities.

Additionally, simplifying network complexity without compromising performance is a key area of interest. This concept involves optimizing the model architecture to balance complexity and efficiency, making it more adaptable and robust for clinical applications.

The integration of region-based CNNs, deep segmentation, and transformer architecture represents a forward-looking approach in the ongoing quest to advance brain tumor classification. These innovative methodologies could significantly contribute to more accurate, reliable, and efficient diagnostic tools in medical imaging.

**Conclusion**

The results of this study underscore the exceptional performance of deep learning models in the classification of brain tumors. Our comparative analysis of various models, including a 3-layer CNN, EfficientNETB4, VGG19, InceptionV3, and U-Net, demonstrates the superior efficacy of the U-Net architecture in this task.

The U-Net model achieved the highest accuracy of 99.5%, an F-score of 98%, a recall of 98.5%, a precision of 98%, and an AUC of 99.8%. These results highlight U-Net's remarkable ability to accurately classify different types of brain tumors, making it a robust tool for medical diagnostics. The 3-layer CNN also performed well, with an accuracy of 97.65%, F-score of 97.65%, recall of 97%, precision of 98%, and an AUC of 99%, showing its viability as a strong alternative.

EfficientNETB4, VGG19, and InceptionV3, although slightly less accurate than U-Net and the 3-layer CNN, still demonstrated significant potential in brain tumor classification tasks. EfficientNETB4 achieved an accuracy of 93%, with an F-score of 92%, recall of 92.5%, precision of 92%, and an AUC of 93%. VGG19 and InceptionV3 also showed competitive results, with accuracies of 90% and 91%, respectively.

The application of these models, particularly U-Net, in brain tumor classification tasks is transformative. The high accuracy and robust performance metrics suggest that these models can be effectively integrated into clinical workflows, aiding in early diagnosis and improving treatment outcomes. The use of transfer learning and data augmentation techniques further enhances the predictive capability of these models, making them reliable tools for healthcare professionals.

In conclusion, the deployment of advanced deep learning models, especially U-Net, represents a significant advancement in the field of medical image classification. These models not only offer high accuracy and reliability but also have the potential to revolutionize the diagnosis and treatment of brain tumors, ultimately improving patient care and outcomes. As AI technology continues to evolve, its integration into medical diagnostics will play an increasingly critical role in advancing healthcare.

**Data availability**

All data for this study were obtained from publicly accessible repositories, so no new experiments involving human subjects were conducted. The analysis relies exclusively on open-access MRI images.